\documentclass[11pt,a4paper]{article}
\usepackage{graphicx}
\usepackage{amsmath,amssymb}
\usepackage{natbib}

\def\indep{\perp\kern-0.50em \perp}

\def\indep{\perp\kern-0.50em \perp}

\begin{document}

\title{
Latent variable models for multivariate dyadic data with zero inflation: Analysis of
intergenerational exchanges of family support}
\author{Jouni Kuha \hspace*{2em}Siliang Zhang \hspace*{2em} Fiona
Steele\\[1ex]
Department of Statistics, London School of Economics and Political
Science}
\date{\today}

\maketitle

\begin{abstract}
Understanding the help and support that is exchanged between family
members of different generations is of increasing importance, with research
questions in sociology and social policy focusing on both predictors of
the levels of help given and received, and on reciprocity between them.
We propose general latent variable models for analysing such data, when
helping tendencies in each direction are measured by multiple binary
indicators of specific types of help. The model combines two continuous
latent variables, which represent the helping tendencies, with two
binary latent class variables which allow for high proportions of
responses where no help of any kind is given or received. This defines a
multivariate version of a zero inflation model. The main part of the
models is estimated using MCMC methods, with a bespoke data augmentation
algorithm. We apply the models to analyse exchanges of help between
adult individuals and their non-coresident parents, using survey data from
the UK Household Longitudinal Study.
\end{abstract}

\vspace*{2ex}

\section{Introduction}
\label{s_intro}

In this article we propose and apply latent variable models for the
joint distribution of variables within a dyad of two interacting units.
This is motivated by research questions in sociology and social policy
about exchanges of help and support between adult individuals and their
non-coresident parents. In all societies such intergenerational
transfers have major implications for individual, family, and societal
wellbeing \citep{mason+lee18}. Transfers between adult children and
their parents are an important element of intergenerational linkages and
a means of providing support to those in need \citep{kunemundetal05},
especially in a context of shrinking social services \citep{pickard15}.
Increases in life expectancy imply an increase in the volume of help
needed by older people with age-related functional limitations.  At the
same time, there may be an increased need for assistance in younger age
groups as a result of delayed transitions to adulthood, precarious
employment, and increasingly diverse and complex family life courses
\citep{lesthaeghe14, henrettaetal18}. Analysis of the factors associated
with exchanges of support between generations is
important in order to anticipate which population sub-groups may be at
risk from lack of support in the future, either because of an increased
unmet need for help or a reduced capacity to provide help among
potential donors.

We consider two broad research questions on such intergenerational
support: what characteristics of individuals and their parents are
associated with different levels of help given and received between
them, and what is the extent and nature of reciprocity of exchanges
(i.e.\ to what extent do children with a high tendency to provide help
to parents  also have a high or low tendency to receive help). Previous
research suggests that reciprocity, either contemporaneous or over the
life course, is an important motivating factor in intergenerational
exchanges of support (e.g.\ \citealt{grundy05};
\citealt{silversteinetal02}).  For example, studies that have analysed
contemporaneous reciprocity have found a positive association between
the provision of support to parents and the receipt of support by their
adult offspring in the U.S.\ (\citealt{chengetal13}) and Britain
(\citealt{grundy05}; \citealt{steele+grundy21}), and studies that have
examined reciprocity across the lifecourse have found that a higher
level of parental support during childhood is associated with an
increased propensity to help parents in later life (e.g.\
\citealt{silversteinetal02}).  Another reason for considering the extent
to which child-parent exchanges are balanced is that being unable to
reciprocate support may have negative consequences for the mental health
and wellbeing of older people (e.g.\ \citealt{davey+eggebeen98}).

Research on intergenerational support is framed by theoretical
perspectives from sociology, social psychology and economics (see e.g.\
the discussions in \citealt{silversteinetal02}, \citealt{grundy05} and
\citealt{kalmijn14}, and references therein). A prominent distinction is
between explanations which focus on altruism and ones which focus on
considerations of the costs and benefits of giving support, although
these motivations do not need to be mutually exclusive. The theories in
turn inform considerations of possible explanatory variables for levels
of support. Many of these variables can be seen as instances of two
broad kinds of factors: \emph{capacity} (financial and time resources)
of the provider of help, and the \emph{needs} of the recipient
\citep{fingermanetal15}. A wide range of such predictors have been
examined for exchanges of support between generations in different
contexts (see the studies cited in this section, and references
therein).

Our goal is to improve the methodology of analysing such questions. We
consider the case of Britain, using cross-sectional survey data from the
Family Network module of the UK Household Longitudinal Study (UKHLS).
These data are described in Section~\ref{s_data}. They include sixteen
questions (`items') about exchanges of help within the dyad defined by a
survey respondent and their non-coresident parent or parents. Eight of
the items indicate whether or not the respondent gives each of eight
specific types of help to the parents (for example, helping them with
housework), and eight indicate types of help that they may receive from
the parents. These items are regarded as measures of two latent
variables, which we interpret as the general tendencies to give and to
receive help. We thus have `doubly multivariate' data, with two sets of
observed binary items measuring two latent variables. The substantive
research questions correspond to questions about the joint distribution
of the latent variables, on their means and on the association between
them.

The analysis of this situation should be handled with an appropriate form of \emph{latent
variable modelling}. Further, the data have two peculiar features which
should be allowed for. First, they display a multivariate form of
\emph{zero inflation}, where the proportion of respondents who give a
zero response to all eight items for a latent variable (i.e.\ no help of
any kind given, or none received) is larger than can be accounted for by
basic models. Second, the signal value of specific types of help may be
different for different types of respondents, especially for men and
women (because of gendered patterns of helping) or for people who live
at longer or shorter distances from their parents (because of different
practicalities of different kinds of help). This can be seen as an
instance of \emph{non-equivalence of measurement} in the items.

We propose a general latent variable modelling framework for the
analysis of such doubly multivariate data on dyads. Its starting point
is a conventional model for two continuous latent variables given
covariates, measured by binary items. Non-equivalence of measurement is
represented by letting the measurement component of the model
depend on some covariates. Zero inflation is allowed for by
supplementing the bivariate continuous latent variable with a bivariate
latent class variable which accounts for the excess of all-zero
responses in one or both of the sets of items. This specification
combines and extends several modelling elements, and draws on the
corresponding literatures (we discuss this further in Section
\ref{ss_models_literature}, after the models have been defined in
Section \ref{ss_models_specification}). Beyond the analysis of exchanges
of help within families, the models could also be applied to comparable
dyadic data elsewhere, such as in applications in family psychology and
organisational behaviour, when variables of interest for the dyads are
latent variables measured with multiple indicators.

We contribute to the literature on intergenerational exchanges of
support by addressing two major methodological limitations of previous
research: the measurement of support given or received, and estimation
of reciprocity of exchanges. Most previous studies using data similar to
those collected in UKHLS have reduced the multivariate data on different
types of exchange in a given direction to a single binary variable
indicating whether any support was given or received. Another approach
has been to analyse the sum score of the items using linear regression
(e.g.\ \citealt{chengetal13}), which implicitly assigns equal weight to
each item and ignores zero inflation. It is common to focus on one
direction of exchange only, for example just support given to elderly
parents (e.g.\ \citealt{silversteinetal02}), or to analyse the receipt
and provision of support separately.  The disadvantage of both
approaches is that they preclude investigation of reciprocity of
exchanges, the importance of which has been widely acknowledged. Among
the few studies that have investigated reciprocity, one approach has
been to treat helping tendencies as categorical, by using first latent
class analysis to identify a typology of exchanges and then modelling
class membership using multinomial regression (\citealt{hoganetal93};
\citealt{chan08}), and another has been to treat them asymmetrically, by
including the receipt of support as a predictor of provision of support,
and vice versa (\citealt{grundy05}; \citealt{chengetal13}). A recent study
by \cite{steele+grundy21} models bidirectional exchanges between adult
children and their parents jointly, interpreting the residual
correlation as a measure of reciprocity, but it treats support given and
received as univariate binary responses.

We estimate the models using a two-step approach where the parameters of
the measurement model for the items given the latent variables are
estimated first, and their values are then held fixed in the second step
where the structural model for the latent variables is estimated. The
second step is carried out using Markov Chain Monte Carlo (MCMC)
methods. Because doing this with general MCMC packages was very slow
(mainly because of the combination of continuous and categorical latent
variables in the models), the estimation was implemented  using a
tailored algorithm written for these models. It has a convenient data
augmentation structure which alternates between sampling the latent
variables given the model parameters and the observed data, and sampling
the parameters given the observed and latent data. These methods of
estimation are described in Section \ref{s_estimation}, with details of
the MCMC algorithm given in Appendix~\ref{s_appendices}.

Our analysis of intergenerational exchanges of help is then described in
Section \ref{s_analysis}. The results show a number of clear
regularities in how the levels of help are associated with different
aspects of the individuals' and their parents' need for help and
capacity to give help. Helping tendencies in the two directions are
positively correlated, conditional on the explanatory variables,
suggesting a substantial amount of reciprocity in helpfulness between
the generations.

\section{Data on exchanges of help between generations}
\label{s_data}

We use data from the UK Household Longitudinal Study (UKHLS), also known
as `Understanding Society' (\citealt{ukhls18}; see \citealt{knies18} for
more information on the study). This is a longitudinal survey of the
members of approximately 40,000 households (at Wave 1 of UKHLS) drawn
from the residential population living in private households in the
United Kingdom. UKHLS started in 2009, but it also subsumes the
smaller British Household Panel Survey (BHPS) which began in 1991.

Information on exchanges of help with parents living outside a
respondent's household was collected in the Family Network
module which was administered in 2001, 06, 11/12, 13/14, and 15/16 (BHSP
Waves 11 and 16 and UKHLS Waves 3, 5, and 7). We carry out
cross-sectional analysis which uses only the last of these, Wave 7 (with
the exception that the development and estimation of the measurement
models was done using pooled data across all five waves; this is
explained separately in Section \ref{ss_analysis_measurement}). In this
module, respondents with at least one non-coresident parent were asked
whether they `nowadays' gave `regularly or frequently' the following
eight types of help to their parent(s), each with a yes--no response:
`giving them lifts in your car (if you have one)' [referred to as
\emph{lifts} below], `shopping for them' (\emph{shopping}), `providing
or cooking meals' (\emph{meals}), `helping with basic personal needs
like dressing, eating or bathing' (\emph{personal care}), `washing,
ironing or cleaning' (\emph{housework}), `dealing with personal affairs
e.g.\ paying bills, writing letters' (\emph{personal affairs}),
`decorating, gardening or house repairs' (\emph{diy}), and `financial
help' (\emph{financial}). The same questions were asked about receipt of
support from the parents, but with the personal care item replaced by
`looking after your children' (\emph{childcare}).

Although respondents were asked to report on giving parents a lift in
their car if they had one, the recorded variable had only `yes' or
`no' responses.  We therefore used other survey information to set this
item to missing for respondents who did not have access to a car.
Similarly, the childcare item was coded as missing for respondents who
did not have dependent children aged 16 or under. For the item on
receiving lifts from parents, we do not have information on whether
the parents have access to a car, so responses of `no' to this item
will include also cases where they do not.

These family network data have several limitations, which are shared by
other large-scale studies of intergenerational exchanges. First, for
practical reasons, data on exchanges with non-coresident relatives were
collected only from the perspective of the survey respondent, in our
case the child. As noted by \citet{chan+ermisch15}, studies with matched
pair data which are collected from both non-coresident parents and
children in the same family are rare and tend to be for small
selective samples. For UKHLS this means that we have rich data on
children, but much less information on their parents. Child responses
may also suffer from reporting bias, for example over-reporting of help
given and under-reporting of help received.  Second, respondents were
asked to report on exchanges with both parents collectively, even when
both are surviving and non-coresident with the respondent, so it is not
possible to distinguish between exchanges with the mother and with the
father, even when they are living apart. Where a respondent had both
biological and step/adoptive parents alive, the recorded responses refer
to the ones that the respondent had most contact with.

A set of covariates was considered to capture factors that may be
associated with help given or received between individuals and their
parents. They are gender, age, partnership status and employment status
of the respondent, the presence and age of children in their family, age
of the oldest non-coresident parent, whether any parent lived alone, and
the travel time between the respondent and the parent living closest.

The sample was first restricted to the 19,052 respondents in UKHLS Wave 7
who were aged 16 or over and who had  at least one non-coresident parent
but no coresident parent. Respondents living with a parent were mainly
younger respondents who had not left the parental home; they were
excluded because their exchanges with their non-coresident parent are
likely to differ from those of respondents who do not live with either
parent. Also excluded were respondents whose closest parent lived or worked abroad
($n$=2590) and those with missing data on any
covariate or on \emph{all} of the help items ($n$=1589). This gives our
main analysis sample of 14,873 respondents. Most of the omissions from
missing data (1226 cases) were due to the indicator of whether either parent
lived alone, while nonresponse on the other covariates and the help
items was much rarer.

The percentages of respondents in the analysis sample who reported giving and
receiving each type of help are shown in Table \ref{t_items}, and
descriptive statistics and coding for the covariates in Table
\ref{t_covariates}. We note that less than half of the respondents
report giving (43.3\%) or receiving (40.0\%) even one of these kinds of
help. This large proportion of all-No responses is a feature that we
will want allow for in modelling these data. The specification and
estimation of the models is described in Sections \ref{s_models} and
\ref{s_estimation}. We will then return to the analysis of the data in
Section \ref{s_analysis}.

\vspace*{3ex}

\begin{table}[h]
\centering
\caption{Percentage of respondents giving help to their non-coresident parents and
receiving help from the parents, by item.}
\label{t_items}
\begin{tabular}{lrr}
\hline
 & Help given & Help received \\
Item & to parents & from parents \\
\hline
Lifts in car (\emph{lifts})                            & 29.2  & 11.2 \\
Shopping (\emph{shopping})                             & 20.7      & 8.0 \\
Providing or cooking meals (\emph{meals})              & 12.3      & 13.9 \\
Basic personal needs (\emph{personal care})            & 3.6       & -- \\
Looking after children (\emph{childcare})              & --        & 40.3 \\
Washing, ironing or clearning (\emph{housework})       & 7.7       & 5.6 \\
Personal affairs (\emph{personal affairs})             & 16.9      & 2.3 \\
Decorating, gardening or house repairs (\emph{diy})    & 17.9      & 8.3 \\
Financial help (\emph{financial})                      & 6.3       & 13.0
\\[2ex]
\emph{At least one of the eight kinds of help:} &43.3& 40.0\\
\hline
\multicolumn{3}{p{0.85\textwidth}}{\footnotesize{Data from UKHLS, Wave
7. Valid
percentages, excluding cases with missing data.
The sample sizes are $n=$14,866 for items on help to parents, and
$n=$14,867 for help from parents. Of these, the \emph{lifts} item is
missing
for the 19.0\% of respondents who have no access to a car, and
\emph{childcare} is missing for the 52.3\% who have no coresident dependent children.}}
\end{tabular}
\end{table}

\begin{table}
\centering
\caption{Descriptive statistics for the covariates used in the analysis
($n$=14,873).}
\label{t_covariates}
\begin{tabular}{lrr}
\hline
Variable &  $n$ & Percent \\
\hline
\textbf{Respondent (child) characteristics}         &           & \\
Age (years)                        & Mean=42.7 & SD=11.4 \\[.5ex]
Gender                                              &           & \\
~~Female                                            & 8570      & 57.6 \\
~~Male                                              & 6303      & 42.4
\\[.5ex]
Partnership status                                  &           & \\
~~Partnered                                         & 11,354    & 76.3 \\
~~Single                                            & 3519      & 23.7
\\[.5ex]
Employment status                                   &           & \\
~~Employed                                          & 11,527    & 77.5 \\
~~Unemployed                                        & 590       & 4.0 \\
~~Economically inactive                             & 2756      & 18.5
\\[.5ex]
%Presence and age of children                        &           & \\
%~~No children                                       & 6022      & 40.5 \\
%~~Any children, of age:&&\\
%\hspace*{2em}$<2$ years                   & 1273      & 8.6 \\
%\hspace*{2em} 2--4 years                  & 2160      & 14.5 \\
%\hspace*{2em} 5--10 years                 & 3675      & 24.7 \\
%\hspace*{2em} 11--16 years                & 3138      & 21.1 \\
%\hspace*{2em} $>16$ years                  & 2911      & 19.6 \\[.5ex]
Age of youngest child                        &           & \\
~~No children                                       & 6022      & 40.5 \\
~~$0-1$ years                   & 1273      & 8.6 \\
~~$2-4$ years                  & 1696      & 11.4 \\
~~$5-10$ years                 & 2306      & 15.5 \\
~~$11-16$ years                & 1724      & 11.6 \\
~~$>16$ years                  & 1852      & 12.5 \\[8pt]
%Age (years)                                         & Mean=42.7 & SD=11.4 \\[8pt]
\textbf{Parent characteristics}                     &           &
\\
Age of oldest parent (years)                        & Mean=71.1 &
SD=11.3 \\[.5ex]
At least one parent lives alone                     &           & \\
~~Yes                                               & 5626      & 37.8 \\
~~No                                                & 9247      & 62.2 \\[8pt]
\textbf{Child-parent characteristics} && \\
Travel time to nearest parent                       &           & \\
~~1 hour or less                                    & 10,771    & 72.4 \\
~~More than 1 hour                                  & 4102      & 27.6 \\
\hline
\end{tabular}
\end{table}

\clearpage

\section{Latent variable models for dyadic data}
\label{s_models}

Here we define, in Section \ref{ss_models_specification}, the latent
variable models that we propose for analysing dyadic data like those
introduced in Section \ref{s_data}. In Section
\ref{ss_models_literature} we discuss how different elements of this
specification draw on previous literature. For ease of exposition the
dyads and variables are mostly introduced with reference to their
meaning in the application to intergenerational help, but we note that
the models are also applicable to any data with a similar structure.

\subsection{Model specification}
\label{ss_models_specification}

Consider data on variables
$(\mathbf{X}_{i},\mathbf{Y}_{Gi},\mathbf{Y}_{Ri})$ for a sample of $n$
dyads $i=1,\dots,n$, where $\mathbf{X}_{i}$ is a $q\times 1$ vector of
covariates and $\mathbf{Y}_{Gi}=(Y_{G1i},\dots,Y_{Gp_{G}i})'$ and
$\mathbf{Y}_{Ri}=(Y_{R1i},\dots,Y_{Rp_{R}i})'$ are two vectors of binary
indicator variables (\emph{items}). In our application, a dyad is
composed of an individual (survey respondent) and their non-coresident parents, and
the items are the questions on the $p_{G}=p_{R}=8$ specific types of
help the respondent gives to the parents ($\mathbf{Y}_{Gi}$) or receives
from the parents ($\mathbf{Y}_{Ri}$). Each item is coded 1 if that kind
of help is given or received, and 0 if it is not. We treat
$\mathbf{Y}_{Gi}$ as multiple indicators of a latent variable
$\eta_{Gi}$ which describes an individual's tendency to give help to
their parents, and $\mathbf{Y}_{Ri}$ as indicators of another latent
variable $\eta_{Ri}$ which describes the parents' tendency to give help
to the individual. We take $\eta_{Gi}$ and $\eta_{Ri}$ to be continuous
variables. The goal is to estimate their joint distribution and how it
depends on the covariates.

Our data contain a substantial number of respondents for whom all of the
items in $\mathbf{Y}_{Gi}$ or $\mathbf{Y}_{Ri}$ are 0 (see Table
\ref{t_items}). The frequencies of such all-zero response patterns are
higher than what would be expected under standard  models with
continuous latent variables. To allow for this, we introduce for each of
the two sets of items a second, binary latent variable.
It defines two latent classes, denoted by 0 and 1, where
class 0 accounts for the excess zeros.
For help to parents, this latent class variable is
denoted $\xi_{Gi}$. The \emph{measurement model} for how $\mathbf{Y}_{Gi}$ measures the latent
variables is then specified by
\begin{eqnarray}
p(\mathbf{Y}_{Gi}=\mathbf{0}|\xi_{Gi}=0,\mathbf{X}_{i})&=&1
\hspace*{2em}\text{ and}
\label{meas1} \\
p(\mathbf{Y}_{Gi}|\xi_{Gi}=1,\eta_{Gi},\mathbf{X}_{i};\boldsymbol{\phi}_{G})&\equiv&
p_{1}(\mathbf{Y}_{Gi}|\eta_{Gi},\mathbf{X}_{i};\boldsymbol{\phi}_{G})
=\prod_{j}p_{1}(Y_{Gji}|\eta_{Gi},\mathbf{Z}_{i};\boldsymbol{\phi}_{G})
\label{meas2}
\end{eqnarray}
where $p(\cdot|\cdot)$ denotes a conditional distribution,
$p_{1}(\cdot|\cdot)$ indicates that a distribution is also conditional
on $\xi_{Gi}=1$, $\mathbf{Z}_{i}$ are a subset of $\mathbf{X}_{i}$, and
$\boldsymbol{\phi}_{G}$ are parameters. In other words, when
$\xi_{Gi}=0$, a respondent is certain to answer `No' to all the items in
$\mathbf{Y}_{Gi}$. When $\xi_{Gi}=1$, the probabilities of the responses
depend on the latent helping tendency $\eta_{Gi}$ and covariates
$\mathbf{Z}_{i}$, and the different items $Y_{Gji}$ are taken to be
conditionally independent of each other; this is a conventional latent
variable model for the binary items, with the extension that the
measurement may be non-equivalent with respect to some covariates
$\mathbf{Z}_{i}$. Together, (\ref{meas1}) and (\ref{meas2}) define a
zero-inflation model where the class $\xi_{Gi}=0$ allows for that part
of the probabilities of $\mathbf{Y}_{Gi}=\mathbf{0}$ which are not
accounted for by the distribution of $\eta_{Gi}$ and the
measurement model given $\eta_{Gi}$.

The measurement model for $\mathbf{Y}_{Ri}$ given
$(\xi_{Ri},\eta_{Ri},\mathbf{Z}_{i})$ is defined similarly, with
parameters $\boldsymbol{\phi}_{R}$. We assume that $\mathbf{Y}_{Gi}$ do
not depend on $(\xi_{Ri},\eta_{Ri})$, $\mathbf{Y}_{Ri}$ do not depend on
$(\xi_{Gi},\eta_{Gi})$, and $\mathbf{Y}_{Gi}$ and $\mathbf{Y}_{Ri}$ are
conditionally independent of each other, and define
$\boldsymbol{\phi}=(\boldsymbol{\phi}_{G},\boldsymbol{\phi}_{R})$. Some
of the items  in $\mathbf{Y}_{Gi}$ and/or $\mathbf{Y}_{Ri}$ may be
missing, in which case the products over $j$ in (\ref{meas2}) and the
corresponding model for $\mathbf{Y}_{Ri}$ are over only those items
which are observed for that respondent. This implies that these missing
data are assumed to be missing at random. We assume here that there are
no missing data in the covariates $\mathbf{X}_{i}$.

The model for the latent variables given the explanatory variables is
specified by the distributions $p(\xi_{Gi}=j,\xi_{Ri}=k|\mathbf{X}_{i};
\boldsymbol{\psi}_{\xi}) \equiv \pi_{jk}(\mathbf{X}_{i};
\boldsymbol{\psi}_{\xi})$ and $p(\eta_{Gi},\eta_{Ri}|\mathbf{X}_{i};
\boldsymbol{\psi}_{\eta})$, where
$\boldsymbol{\psi}=(\boldsymbol{\psi}_{\xi},\boldsymbol{\psi}_{\eta})$
are parameters, and $(\eta_{Gi},\eta_{Ri})$ and $(\xi_{Gi},\xi_{Ri})$ are
taken to be independent of each other given $\mathbf{X}_{i}$. We refer to this as the
\emph{structural model} for the latent variables. It will be the focus
of interest for substantive research questions about help between
individuals and their parents.

Let $\mathbf{Y}=(\mathbf{Y}_{G},\mathbf{Y}_{R})$ denote all of the
observed data on $\mathbf{Y}_{i}=(\mathbf{Y}_{Gi},\mathbf{Y}_{Ri})$, and
$\mathbf{X}$ all the $\mathbf{X}_{i}$. Define $G_{i}=1$ if $\mathbf{Y}_{Gi}\ne
\mathbf{0}$ and $G_{i}=0$ if $\mathbf{Y}_{Gi}=\mathbf{0}$, and define
$R_{i}$ similarly for $\mathbf{Y}_{Ri}$.
If we take the observations $i$ to be independent,
the log-likelihood for the model is
\begin{eqnarray}
\lefteqn{\log p(\mathbf{Y}|\mathbf{X};\boldsymbol{\phi},\boldsymbol{\psi})} \nonumber \\
&\hspace*{-1em}=& \hspace*{-1em}
\sum_{i=1}^{n}
\log\left[
\pi_{11}(\mathbf{X}_{i};\boldsymbol{\psi}_{\xi})\,\int\!\!\int
p_{1}(\mathbf{Y}_{Gi}|\eta_{Gi},\mathbf{Z}_{i};\boldsymbol{\phi}_{G})
\,p_{1}(\mathbf{Y}_{Ri}|\eta_{Ri},\mathbf{Z}_{i};\boldsymbol{\phi}_{R})
\,p(\eta_{Gi},\eta_{Ri}|\mathbf{X}_{i};\boldsymbol{\psi}_{\eta})
\,d\eta_{Gi}d\eta_{Ri}
\right. \nonumber  \\
&&\hspace*{3em} + (1-R_{i})\;
\pi_{10}(\mathbf{X}_{i};\boldsymbol{\psi}_{\xi})\,\int
p_{1}(\mathbf{Y}_{Gi}|\eta_{Gi},\mathbf{Z}_{i};\boldsymbol{\phi}_{G})
\;p(\eta_{Gi}|\mathbf{X}_{i};\boldsymbol{\psi}_{\eta})
\,d\eta_{Gi}
 \nonumber  \\
&&\hspace*{3em} + (1-G_{i})\;
\pi_{01}(\mathbf{X}_{i};\boldsymbol{\psi}_{\xi})\,\int
p_{1}(\mathbf{Y}_{Ri}|\eta_{Ri},\mathbf{Z}_{i};\boldsymbol{\phi}_{R})
\,p(\eta_{Ri}|\mathbf{X}_{i};\boldsymbol{\psi}_{\eta})
\,d\eta_{Ri}
 \nonumber  \\
&&\hspace*{3em} \left.
\vphantom{\int p_{1}(\mathbf{Y}_{Gi}|\eta_{Gi},\mathbf{Z}_{i};\boldsymbol{\phi}_{G})}
+(1-G_{i})(1-R_{i}) \;
\pi_{00}(\mathbf{X}_{i};\boldsymbol{\psi}_{\xi})\right].
\label{loglik}
\end{eqnarray}
We further specify the structural model for each $i=1,\dots,n$ as
\begin{equation}
p(\eta_{Gi},\eta_{Ri}|\mathbf{X}_{i}; \boldsymbol{\psi}_{\eta}) \sim N
\left(
\begin{bmatrix}
% no delimiters; bmatrix: [; pmatrix: (; vmatrix: |
\boldsymbol{\beta}_{G}'\mathbf{X}_{i} \\
\boldsymbol{\beta}_{R}'\mathbf{X}_{i}
\end{bmatrix}
, \;
\begin{bmatrix}
\sigma^{2}_{G} & \\
\rho_{GR}\sigma_{G}\sigma_{R} & \sigma^{2}_{R}
\end{bmatrix}
\right)
\label{structural1}
\end{equation}
and
\begin{equation}
\log \left[
\frac{\pi_{jk}(\mathbf{X}_{i};
\boldsymbol{\psi}_{\xi})}{\pi_{00}(\mathbf{X}_{i}; \boldsymbol{\psi}_{\xi})}\right] =
\boldsymbol{\gamma}'_{jk}\mathbf{X}_{i}
\label{structural2}
\end{equation}
for $j,k=0,1$ with $\boldsymbol{\gamma}_{00}=\mathbf{0}$, i.e.\ as a
bivariate normal linear model for $(\eta_{Gi},\eta_{Ri})$ and a
multinomial logistic model for $(\xi_{Gi},\xi_{Ri})$.
Thus here
$\boldsymbol{\psi}_{\eta}$ includes
$(\boldsymbol{\beta}_{G},\boldsymbol{\beta}_{R},\sigma^{2}_{G},\sigma^{2}_{R},\rho_{GR})$ in (\ref{structural1}), and
$\boldsymbol{\psi}_{\xi}$ includes
($\boldsymbol{\gamma}_{01}, \boldsymbol{\gamma}_{10},
\boldsymbol{\gamma}_{11}$) in (\ref{structural2}). Finally,
the measurement models given the continuous latent variables are
specified as
\begin{equation}
\text{logit}[p_{1}(Y_{Gji}=1|\eta_{Gi},\mathbf{Z}_{i};\boldsymbol{\phi}_{G})]=
\tau_{Gj} +  \boldsymbol{\delta}_{Gj}'\mathbf{Z}_{i}
+(\lambda_{Gj}
+ \boldsymbol{\zeta}_{Gj}'\mathbf{Z}_{i})\, \eta_{Gi}
\label{meas_YT}
\end{equation}
for $j=1,\dots,p_{G}$, so that $\boldsymbol{\phi}_{G}$ consists of all
the $\tau$, $\boldsymbol{\delta}$, $\lambda$ and $\boldsymbol{\zeta}$
parameters for $\mathbf{Y}_{Gi}$, and the models for the items
in $\mathbf{Y}_{Ri}$ are specified similarly, with parameters $\boldsymbol{\phi}_{R}$.
The baseline parameters of these models are the intercepts (the $\tau$s)
and the loadings of the latent $\eta$ variables (the $\lambda$s). These
are then further modified by the covariates if any of the
$\boldsymbol{\delta}$ or $\boldsymbol{\zeta}$ parameters are non-zero,
in which case the measurement model for that item is non-equivalent with
respect to the corresponding variables in $\mathbf{Z}_{i}$. For simplicity, we consider
only models where any non-equivalence in an item affects both the
intercept and the loading, so that for that item the elements of
$\boldsymbol{\delta}$ and $\boldsymbol{\zeta}$ corresponding to the same
variable in $\mathbf{Z}_{i}$ are either both zero or both
non-zero. The motivation and choice of $\mathbf{Z}_{i}$ in our
application are discussed in Section \ref{ss_analysis_measurement}.

\subsection{Previous literature on the elements of the models}
\label{ss_models_literature}

The model specified in Section \ref{ss_models_specification} combines
several existing modelling elements, and draws on the corresponding
literatures. The starting point is the conventional general framework
for latent variable modelling with covariates (see e.g.\
\citealt{skrondal+rabehesketh04} and \citealt{bartholomewetal11}). If
$(\eta_{G},\eta_{R})$ were the only latent variables, this would be a
standard model for the joint distribution of two continuous latent
variables given covariates $\mathbf{X}$. When, as here, all the measures
$\mathbf{Y}$ of the latent variables are binary and the measurement
models for them are logistic models, this is a common instance of
what is known, especially in psychometrics and educational testing, as
Item Response Theory (IRT) modelling (see e.g.\ \citealt{deayala09} and
\citealt{vanderlinden16}).

Including covariates $\mathbf{Z}$ in a measurement model, as we do in
(\ref{meas_YT}), allows the measurement of a latent variable to be
non-equivalent with respect to these covariates. This is also a standard
approach in applications where such non-equivalence may be of concern,
such as cross-national survey research and other `multigroup'
situations, as well as many applications of IRT, where non-equivalence
of measurement is commonly known as differential item functioning (DIF).
For overviews of these ideas and methods, see \cite{kankarasetal11b} and
\cite{millsap11}.

The least familiar element of the model is the way we allow for the
large number of all-zero responses by adding the latent class
variables $(\xi_{R},\xi_{R})$. To motivate this, consider first models
for a single non-negative variable $Y$ with excess zeros, meaning that
the observed probability $P(Y=0)$ is greater than
can be expected under an assumed distribution $p(Y)$ for $Y$. There are,
broadly, three ways of representing this situation, depending on how
many of the zero values are taken to be accounted for by $p(Y)$: (1) all
of them---\emph{censoring} models where it is assumed that $Y$ could
actually be negative but that all such values are recorded as 0, so that
$P(Y=0)=p(Y\le 0)$; (2) none of them---\emph{hurdle} models where we
model separately $P(Y=0)$ and $p(Y|Y>0)$; or (3) some of
them---\emph{zero-inflated} models where $P(Y=0)=\pi+(1-\pi)p(Y=0)$ with
an additional probability parameter $\pi$ for the proportion of zeros
which is not accounted for by $p(Y)$ (see e.g.\ \citealt{tobin58},
\citealt{cragg71}, \citealt{mullahy86}, \citealt{lambert92}, and
\citealt{min+agresti05} for introductions and comparisons of these
possibilities).

We are interested in latent-variable models for multivariate items.
Denote for the moment a generic continuous latent variable by $\eta$,
and its indicators by $\mathbf{Y}$, omitting covariates, so that the
model is specified by $p(\mathbf{Y}|\eta)p(\eta)$, and suppose that the
observed proportion of $\mathbf{Y}=\mathbf{0}$ is so high that we want to
allow for it specially. Here the basic model is in effect already a
censoring model, in that estimates of its parameters will be determined
so that they accommodate these zeros. This, however, can distort the
parameters in a way which leaves the model as a whole badly specified to
account for the non-zero patterns of responses (see \citealt{walletal15}
for a discussion of the biases which can arise when a latent-variable
model is poorly specified in this way). A hurdle model is also
unappealing here, because it would involve conditioning on the observed
items $\mathbf{Y}$. This leads us to consider zero-inflated models,
extended to multivariate $\mathbf{Y}$.

These models can be seen as an instance of \emph{finite mixture models}.
The general form of them is here $\sum_{g}
p_{g}(\mathbf{Y}|\eta)\,p_{g}(\eta)\,\pi_{g}$, where $\pi_{g}=P(\xi=g)$
are probabilities of a latent-class variable~$\xi$. One type
of such models is obtained when
$p_{g}(\mathbf{Y}|\eta)=p(\mathbf{Y}|\eta)$, i.e.\ when the measurement
model is the same in every class~$g$. Then the model becomes
$p(\mathbf{Y}|\eta)\,p^{*}(\eta)$ where $p^{*}(\eta)=\sum_{g}
p_{g}(\eta)\,\pi_{g}$ is a finite mixture distribution. This provides a
way of specifying the basic latent variable model with a more flexible
distribution for $\eta$ than is possible with a single parametric (for
example normal) distribution. Mixture modelling with this purpose is
discussed by \citet{walletal12} and \citet{walletal15}.
Here, in contrast, we particularly need models where the
measurement models do depend on the class. This represents a situation
where the latent classes correspond to individuals with different
\emph{response styles}, i.e.\ different relationships between the latent
variable $\eta$ and its measures $\mathbf{Y}$. This idea has been used
in various contexts of measurement, especially in applications of
psychological and educational testing; see \citet{walletal15},
\citet{huang16}, and references therein.

A zero-inflation model for multivariate $\mathbf{Y}$ involves two
response styles: one where $\mathbf{Y}=\mathbf{0}$ always, and one where
$\mathbf{Y}$ follows an IRT model given $\eta$. This has been proposed
for models where the items are binary (\citealt{muthen+asparouhov06};
\citealt{finkelmanetal11}; \citealt{walletal15}), ordinal
\citep{magnus+liu18} or count variables \citep{magnus+thissen17},
sometimes with extensions such as separate classes for all-0 and all-1
response patterns or more than one class for general response patterns.
Our model is similar to the previous ones for binary items (although
different versions use ostensibly different but equivalent
specifications for $\eta$ in the all-zero class for $\xi$). To
accommodate the dyadic data, however, we have extended them to include
two latent variables ($\eta_{G}$ and $\eta_{R}$), with separate
zero-inflation classes for each of them.

\section{Estimation of the models}
\label{s_estimation}

\subsection{Two-step estimation}
\label{ss_estimation_twostep}

We employ a two-step approach to  estimate these models.
What this means is that the measurement model is first selected and
estimated separately, and its parameters $\boldsymbol{\phi}$ are then
fixed at their estimated values for all subsequent exploration and
estimation of the structural model. These two steps for our models are
described separately in Sections \ref{ss_estimation_measurement} and
\ref{ss_estimation_structural} below.

This idea of two-step estimation of latent variable models goes back to \citeauthor{burt73}
(\citeyear{burt76}, \citeyear{burt73}), and the theory and application
of it have been developed more recently by \citet{xue+bandeen-roche02}
and \citet{bakk+kuha18}. Our motivation for using it here is twofold.
First, it substantially reduces the computational demands compared to
the `one-step' method of estimating all parts of the models together.
This is beneficial here, where the estimation of even the structural
model on its own is demanding. Second, and as importantly, the
two-step approach means that the measurement models --- and thus the
exact definition of the latent variables --- remain fixed in subsequent
analyses, and do not change when the specification of the structural
model is changed, for example when covariates are added or removed. In
our work, this extends also to other analyses of intergenerational
exchanges of family support outside this paper, where we also want to
keep the definitions of the latent variables fixed in this sense.

\subsection{Estimation of the measurement models}
\label{ss_estimation_measurement}

In the first step of the estimation, the measurement models
 for $\mathbf{Y}_{G}$ and $\mathbf{Y}_{R}$ are estimated separately and conditional on
$\mathbf{Z}$ alone. This means that for $\mathbf{Y}_{G}$ we consider the
log likelihood
\begin{eqnarray}
\log
p(\mathbf{Y}_{G}|\mathbf{Z};\boldsymbol{\phi}_{G},\boldsymbol{\psi}_{G}^{*})
&=&
\sum_{i=1}^{n}
\log\left[
\pi_{G}(\mathbf{Z}_{i}; \boldsymbol{\psi}^{*}_{G\xi})\,\int
p_{1}(\mathbf{Y}_{Gi}|\eta_{Gi},\mathbf{Z}_{i};\boldsymbol{\phi}_{G})
\,p(\eta_{Gi}|\mathbf{Z}_{i};\boldsymbol{\psi}^{*}_{G\eta})
\,d\eta_{Gi}
\right. \nonumber \\
&& \hspace*{4em}
\left.  \vphantom{\int p_{1}(\mathbf{Y}_{Gi}|\eta_{Gi},\mathbf{Z}_{i};\boldsymbol{\phi}_{G})}
+ (1-G_{i}) \,
\left(1-\pi_{G}(\mathbf{Z}_{i};\boldsymbol{\psi}^{*}_{G\xi})\right)
\right]
\label{loglik1}
\end{eqnarray}
where
$p_{1}(\mathbf{Y}_{Gi}|\eta_{Gi},\mathbf{Z}_{i};\boldsymbol{\phi}_{G})$
is as before. The structural model here consists of
$\pi_{G}(\mathbf{Z}_{i};\boldsymbol{\psi}^{*}_{G\xi})
=P(\xi_{i}=1|\mathbf{Z}_{i};\boldsymbol{\psi}^{*}_{G\xi})$ and
$p(\eta_{Gi}|\mathbf{Z}_{i};\boldsymbol{\psi}^{*}_{G\eta})$, specified
as a binary logistic and a normal linear model, and
$\boldsymbol{\psi}_{G}^{*}=
(\boldsymbol{\psi}_{G\xi}^{*},\boldsymbol{\psi}_{G\eta}^{*})$ are the
parameters of these models. Here (\ref{loglik1}) is obtained by
integrating (\ref{loglik}) over
$p(\mathbf{Y}_{R},\mathbf{X}_{*}|\mathbf{Z})$, where $\mathbf{X}_{*}$
denotes the variables in $\mathbf{X}$ but not in $\mathbf{Z}$. This is
actually only approximately true, because if
(\ref{structural1})--(\ref{structural2}) holds given $\mathbf{X}$, then
the structural models given $\mathbf{Z}$ only are generally not exactly
of binary logistic and normal linear form. We ignore this small approximation
and maximize (\ref{loglik1}) to estimate $\boldsymbol{\phi}_{G}$.
The parameters $\boldsymbol{\phi}_{R}$ are estimated similarly from a model
like (\ref{loglik1}) for $\mathbf{Y}_{R}$. Denote these estimates by
$\tilde{\boldsymbol{\phi}}=(\tilde{\boldsymbol{\phi}}_{G},\tilde{\boldsymbol{\phi}}_{R})$.
The estimates of $\boldsymbol{\psi}_{G}^{*}$ and $\boldsymbol{\psi}_{R}^{*}$
from this step are discarded.

\subsection{Estimation of the structural models}
\label{ss_estimation_structural}

In the second step of estimation, the structural models are
then estimated, treating the estimated measurement parameters
$\tilde{\boldsymbol{\phi}}$ from the first step as known numbers. In
other words, the log-likelihood for the second step is (\ref{loglik})
but in the form $\log
p(\mathbf{Y}|\mathbf{X};\tilde{\boldsymbol{\phi}},\boldsymbol{\psi})$
where only $\boldsymbol{\psi}$ are unknown parameters. We omit below the
fixed $\tilde{\boldsymbol{\phi}}$ from the notation for simplicity. We
further write $\boldsymbol{\zeta}=(\boldsymbol{\xi},\boldsymbol{\eta})$,
where $\boldsymbol{\xi}$ denotes all the values of the latent
$(\xi_{Gi},\xi_{Ri})$ for the dyads $i$ in the sample, and
$\boldsymbol{\eta}$ all the values of $(\eta_{Gi},\eta_{Ri})$.

In our analyses, this step was carried out in the Bayesian
framework and using MCMC methods of estimation. The
estimation algorithm has a data augmentation
structure, which alternates between sampling the
latent variables and sampling the model parameters:
\begin{itemize}
\item
\textbf{Imputation step}:
Given the observed data $(\mathbf{Y},\mathbf{X})$ and the most recently sampled value of
the parameters $\boldsymbol{\psi}$, sample a value for the
latent variables $\boldsymbol{\zeta}$ from the conditional distribution
\[
p(\boldsymbol{\zeta}|\mathbf{Y},\mathbf{X},\boldsymbol{\psi})
\propto
p(\mathbf{Y}|\boldsymbol{\zeta},\mathbf{X})p(\boldsymbol{\zeta}|\mathbf{X};\boldsymbol{\psi}).
\]
This is further split into sampling
$\boldsymbol{\xi}$ from
$p(\boldsymbol{\xi}|\boldsymbol{\eta},\mathbf{Y},\mathbf{X},\boldsymbol{\psi})$,
using $\boldsymbol{\eta}$ from the previous iteration,
and then $\boldsymbol{\eta}$ from
$p(\boldsymbol{\eta}|\boldsymbol{\xi},\mathbf{Y},\mathbf{X},\boldsymbol{\psi})$.
\item
\textbf{Posterior step}: Given the observed data
$(\mathbf{Y},\mathbf{X})$ and the most recently sampled value of
the latent variables $\boldsymbol{\zeta}$,
sample a value for the
parameters $\boldsymbol{\psi}$ from the conditional distribution
\[
p(\boldsymbol{\psi}|\mathbf{Y},\mathbf{X},\boldsymbol{\zeta})
=p(\boldsymbol{\psi}|\mathbf{X},\boldsymbol{\zeta})
\propto
p(\boldsymbol{\zeta}|\mathbf{X};\boldsymbol{\psi}) p(\boldsymbol{\psi})
\]
where $p(\boldsymbol{\psi})=p(\boldsymbol{\psi}_{\xi})p(\boldsymbol{\psi}_{\eta})$
is the prior distribution of the parameters, taking
$\boldsymbol{\psi}_{\xi}$
and $\boldsymbol{\psi}_{\eta}$ to be independent a priori.
The conditional
distribution then further splits into
\[
p(\boldsymbol{\psi}|\mathbf{X},\boldsymbol{\zeta})
=p(\boldsymbol{\psi}_{\xi}|\mathbf{X},\boldsymbol{\xi})
\, p(\boldsymbol{\psi}_{\eta}|\mathbf{X},\boldsymbol{\eta})
\propto
[p(\boldsymbol{\xi}|\mathbf{X};\boldsymbol{\psi}_{\xi})
p(\boldsymbol{\psi}_{\xi})]\,
[p(\boldsymbol{\eta}|\mathbf{X};\boldsymbol{\psi}_{\eta})
p(\boldsymbol{\psi}_{\eta})],
\]
which can be sampled separately and in parallel for
$\boldsymbol{\psi}_{\xi}$ and $\boldsymbol{\psi}_{\eta}$. This does not
depend on the measurement items $\mathbf{Y}$, because they are not in
the `Markov blanket' of $\boldsymbol{\psi}$ (in the directed acyclic
graph for the model, $\mathbf{Y}$ are not parents, children or
co-parents of children of $\boldsymbol{\psi}$). The posterior step thus
involves sampling the parameters of two regression models given
$\mathbf{X}$, a multinomial logistic model for $\boldsymbol{\xi}$ and a
bivariate linear model for $\boldsymbol{\eta}$, exactly as if from their
posterior distributions if the most recently imputed values of
$\boldsymbol{\xi}$ and $\boldsymbol{\eta}$ were real observed data.
\end{itemize}

We note that since the measurement models are fixed, the structural
model is straightforwardly identified here. In particular, `label
switching', where the numbering of the latent classes changes between
MCMC iterations, cannot occur.

The details of these steps are described in Appendix
\ref{s_appendices}. We wrote bespoke code to implement them, because the
implementation using general MCMC packages was slow; this was mainly due
to the sampling of the categorical latent variables $\boldsymbol{\xi}$.

\section{Analysis of intergenerational exchanges of help}
\label{s_analysis}

\subsection{Measurement models}
\label{ss_analysis_measurement}

We use measurement models which are non-equivalent with respect to two
covariates~$\mathbf{Z}$: the gender of the respondent, and the distance
between where the respondent and their parents live. It is substantively
to be expected, and empirically confirmed in these data, that the
patterns of what kinds of help a person gives and receives may vary by
these covariates. Some types of help are strongly gendered among the
generations considered here, with men and women expressing their
helpfulness in different ways and receiving different kinds of help from
their parents. Similarly, for obvious practical reasons a longer
distance between the parties may affect some types of help more than
others. This being the case, the expected levels and patterns of
different kinds of help may be different between men and women and
between respondents at different distances from their parents, even for
individuals who actually have a similar latent tendency to give or
receive help. The non-equivalent measurement models allow for this
possibility. We thus define $\mathbf{Z}_{i}=(Z_{gi},Z_{di})$, where
$Z_{gi}$ is an indicator variable for a female respondent and $Z_{di}$
an indicator for a respondent who lives at a distance of  an hour or
more's travel time from their parents.

The measurement models were estimated using more of the UKHLS data than
are used for the second step of estimation discussed in Section
\ref{ss_analysis_structural} below. This was because these models were
intended for use in multiple analyses of the data from the Family
Networks module, and were developed prior to the analysis described
here. The models were first explored for data from the 2001 Wave 11 of
BHPS, to identify items for which non-equivalence with respect to gender
and/or distance was substantial enough that it should be allowed for.
This selection was done using a combination of likelihood ratio tests
and the AIC and BIC statistics. The selected models were then
re-estimated using pooled data from all the five available waves of
UKHLS/BHPS, to maximize the amount of data which contributed to these
estimates (longitudinal observations for a respondent in these pooled
data are not independent; ignoring this, however, affects only the
standard errors of the parameter estimates, which are not needed for
what follows). The models were estimated using maximum likelihood
estimation, with the Mplus 6.12 software \citep{muthen+muthen10}.

The selected measurement models include some non-equivalence in most
items, especially with respect to distance. Of the items on help to
parents ($\mathbf{Y}_{G}$), \emph{financial}, \emph{lifts} and
\emph{diy} are non-equivalent with respect to gender, and all but
\emph{personal affairs}, \emph{personal care} and \emph{financial help}
with respect to distance. Of the items on help from parents
($\mathbf{Y}_{R}$), \emph{financial} and \emph{meals} are non-equivalent
with respect to gender, and all but \emph{personal affairs} and
\emph{financial} with respect to distance. The intercept and loading
parameters of \emph{personal care}, which was fully equivalent for both
$\eta_{G}$ and $\eta_{R}$, were fixed at 0 and 1 respectively in both
measurement models, to fix the measurement scales of the two latent
variables. For each of $\eta_{G}$ and $\eta_{R}$ at least two items
which measure them are equivalent with respect to gender, and at least
two with respect to distance. This means that the structural models for
$(\eta_{G}, \eta_{R})$ given gender and distance are also identified,
separately from the measurement models. However, information that is
available for estimating these associations is clearly reduced,
especially for distance, when the non-equivalent measurement models
account for much of the observed association between distance and the
items $\mathbf{Y}_{G}$ and $\mathbf{Y}_{R}$. The associations between
other variables in $\mathbf{X}$ and $(\eta_{G},\eta_{R})$ in the
structural model are then conditional on gender and distance in this
sense, i.e.\ they refer to the latent variables as they are defined by
these measurement models with this adjustment for non-equivalence.

\begin{figure}[t]
\caption{Item response curves for
the item \emph{diy}
(decorating, gardening or house repairs)
for help that respondents give to their parents, for
probability of giving help conditional on the latent variable
$\eta_{G}$ (tendency to give help), separately for the combinations of gender of the respondent
and distance between respondent and their parents.
The dotted vertical line is
approximate mean of $\eta_{G}$.}
\begin{center}
\includegraphics[height=9cm]{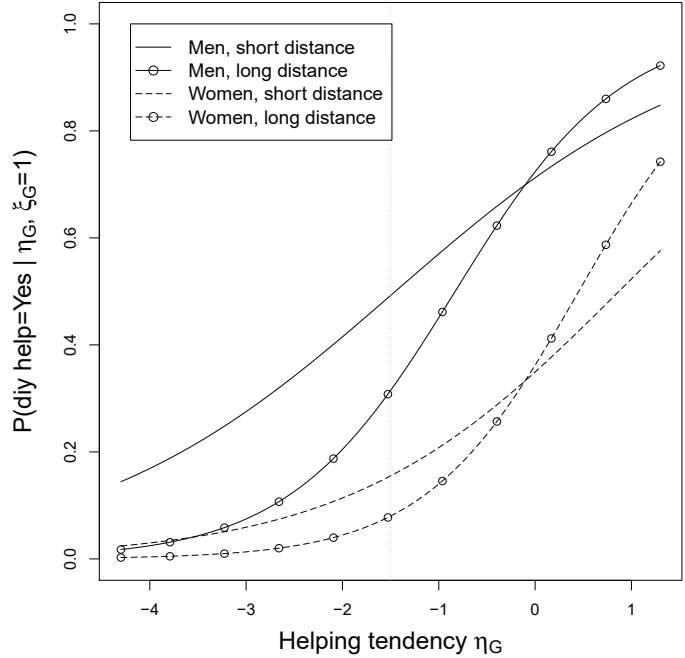}
\end{center}
\label{f_meas1}
\end{figure}

An illustrative example of the estimated measurement models is shown in
Figure \ref{f_meas1}, for the \emph{diy} item (decorating, gardening or
house repairs) on help given to the parents. The model for this item is
non-equivalent with respect to both covariates. The plot shows the
estimated probabilities of giving such help as a function of the latent
tendency of helpfulness $\eta_{G}$, separately for each combination of
gender and distance. Considering the genders, it can be seen that, at
the same level of this tendency, men are more likely to give this
kind of help. The non-equivalence with respect to distance
shows most clearly in the loading (or `discrimination') parameters.
These are larger --- and the probability curves thus steeper --- when a
respondent lives further away from their parents, so that giving such
help is a more discriminating signal of helpfulness for such respondents
than for those who live near their parents.

\subsection{Models for help between respondents and their parents}
\label{ss_analysis_structural}

Fixing the parameters of the measurement models at their estimated values from Section
\ref{ss_analysis_measurement}, we then estimated the structural models
which are our focus of interest. The MCMC algorithm described in Section
\ref{ss_estimation_structural} and the Appendix was run for two MCMC
chains of 110,000 iterations each, from different starting values.
Discarding the first 10,000 iterations of each, conventional convergence
diagnostics indicated that the chains had converged. The two chains were
then combined, so the estimates below are based on 200,000 draws from
the posterior distributions of the parameters.

The estimated parameters of the bivariate linear model
(\ref{structural1}) for the continuous latent variables
$(\eta_{G},\eta_{R})$ are shown in Table \ref{t_model_estimates}. The
coefficients of the multinomial logistic model (\ref{structural2}) for
the categorical latent variables $(\xi_{G},\xi_{R})$ are less convenient
for interpretation, because they express comparisons of the
probabilities in the joint distribution of the variables, relative to
the case $(\xi_{G},\xi_{R})=(0,0)$.
Instead, in Table~\ref{t_pi} we summarise this
model with a focus on the marginal distributions of $\xi_{G}$ and
$\xi_{R}$, using comparisons of fitted probabilities. We first
calculated the probabilities
$p(\xi_{G}=j,\xi_{R}=k|\mathbf{X}_{i};\boldsymbol{\psi}_{\xi})$ for
$j,k=0,1$ given selected covariate values $\mathbf{X}_{i}$ for each of
the $n=14,873$ respondents $i$ and for each of the 200,000 MCMC draws of
the parameters~$\boldsymbol{\psi}_{\xi}$. Table \ref{t_pi} shows these
fitted probabilities, averaged over both respondents and parameter
draws. It also shows the odd ratios between $\xi_{G}$ and $\xi_{R}$
calculated from these averages, as well as the average marginal
probabilities $p(\xi_{G}=1)$ and $p(\xi_{R}=1)$ that a respondent
belongs to the class 1 where they may give and receive help
respectively. Different choices are considered for the $n$ values
of~$\mathbf{X}_{i}$. In the first row of the table, these are the actual
values for the respondents in the observed sample. On the other rows,
one covariate in turn is fixed at a single value for every respondent,
while the other covariates are left at their sample values. For example,
the second row of the table shows the results for a hypothetical sample
where every respondent is aged 35. For cases with different fixed values
of the same covariate, the table also shows the differences of the
marginal probabilities between them, and posterior standard deviations
of these differences over the parameter draws.

As discussed in Section \ref{s_models}, we interpret
$(\eta_{G},\eta_{R})$ as continuous latent tendencies to give and to
receive help. Although $\xi_{G}$ and of $\xi_{R}$ were introduced
primarily to account for zero inflation, on the face of it
they can also be interpreted in terms of binary helping tendencies, with class 0
of each being the class of firm `non-givers' or `non-receivers' of help.
In this sense we can interpret both higher conditional means of
$\eta_{G}$ and $\eta_{R}$, and higher conditional probabilities of class
1 of $\xi_{G}$ and of $\xi_{R}$,
as indications of higher levels of helpfulness of the respondent to the
parents or vice versa; here we refer to both of these as `positive
associations' between a covariate and helpfulness. We note first that
the average marginal probability of class 0 is here 0.27 for $\xi_{G}$
and 0.32 for $\xi_{R}$. Each of these accounts for about half of the
proportions of all-zero responses to the corresponding items (which were
0.57 and 0.60 respectively, as shown in Table \ref{t_items}).

Considering first the models for help given by respondents to the
parents, covariates which are strongly and positively associated with it
in the linear model for $\eta_{G}$ are higher age of the oldest parent,
at least one parent living alone, and the respondent being single or not
employed. Respondents who have no young children at home (i.e.\ have no
children, or only older children) also tend to help more, although this
association is less clear. In the model for $\xi_{G}$, significant
positive associations are also found for age of oldest
parent and a parent living alone, and additionally for younger age of the
respondent. Considering then help received from parents, characteristics
which are positively associated with it in the model
for $\eta_{R}$ are the parents not living alone and the respondent being
younger, single, not employed or having no children at home. Similar
associations are seen in the model for  $\xi_{R}$ for younger and single
respondents, and, in addition, there is a negative association between
help from parents and the respondent having only older children at home.

For help received from parents, associations with
$\eta_{R}$ and $\xi_{R}$ are in different directions for employment
status and for having no children at home. Respondents who are
unemployed or economically inactive, rather than employed, are more
likely to be in the no-help-received class $\xi_{R}=0$ but, if they
are not in this class, the level of help they do receive ($\eta_{R}$)
tends to be higher. Similarly, respondents with no children at home have
higher probability of $\xi_{R}=0$, but otherwise tend to receive more
help. These diverging findings for the categorical and continuous parts of
the model are intriguing, but we hesitate to offer firm substantive
interpretations for them.

In the discussion above, we omitted comments about associations
involving gender of the respondent and the distance between them and the
nearest parent. As discussed in Section \ref{ss_estimation_measurement},
the interpretation for these covariates is somewhat different because
they are also included in the measurement models for the non-equivalent
items in $\mathbf{Y}_{G}$ and $\mathbf{Y}_{R}$. This means, in effect,
that the estimated associations for gender and distance in Tables
\ref{t_model_estimates} and \ref{t_pi} are informed only by those items
which are equivalent with respect to them. Even so, these associations
are strong for $\eta_{G}$ and $\eta_{R}$, with women tending to both
give and to receive more help than men, and the level of help in both
directions being lower when children and parents live far apart.

How, then, should we summarise these results? One way to do so is to
think of the covariates as different instances of two broad categories
of characteristics: an actor's (the child's or the parents')
\emph{capacity to give} help, and the other actor's \emph{need to
receive} help. Considering the models for help to parents in this light,
there is a clear positive association between the parents' need and help
given: older parents and ones who live alone tend to receive more help
from their children. In terms of the children's capacity to help, the
respondent characteristics which are positively associated with helping
(being single and not having young children at home) can perhaps be
interpreted in these terms, if we take `capacity' to mean `opportunity'
in the sense of having fewer other commitments. Conversely, in the
models for help received by the respondents, the negative association of
helpfulness with the parent(s) living alone perhaps reflects their
reduced capacity to help, while the positive associations with the child
being younger, single or not employed may be interpreted as instances of
higher need by the child. We also observe a reduction of helping
behaviour for respondents who live with a partner and/or have young
children at home. Other things being equal, such respondents are less
likely both to give help to their non-coresident parents and to receive
help from them. We might perhaps think of this situation as one of a
self-contained family unit whose support activities may be more
likely to be directed within the family rather than outside of it.

The models also give estimates of the associations between the levels of
help in the two directions, allowing us to examine reciprocity of
support between children and their parents. Here these estimated
associations are strong. For the categorical part of the model, the odds
ratios between $\xi_{G}$ and $\xi_{R}$ (which depend on the covariates)
are typically between 5 and 10. For the continuous part, the conditional
correlation between $\eta_{G}$ and $\eta_{R}$ is 0.50. This is in fact
substantially higher than their marginal correlation, estimated from a
model without covariates (not shown here), which is 0.23. This
difference is mainly due to the age variables. The ages of the
respondent and their oldest parent are strongly associated, with a
sample correlation of 0.87. If we include either one of them alone in
the models, the conditional correlation between $\eta_{G}$ and
$\eta_{R}$ is already about 0.50, and the age variable has a strong
positive association with help given and a negative one with help
received (the models where both are included, as in Tables
\ref{t_model_estimates} and \ref{t_pi}, further indicate, more
specifically, that older respondents tend to receive less help, and
older parents tend to receive more help). The two age effects thus naturally
pull in different directions, so that the marginal correlation of help
given and received is somewhat suppressed. If, however, we condition on the ages,
i.e.\ account for the different levels of help we would expect on
average from children and parents of given ages, the correlation
between them is substantially higher. In this sense, the results
suggest a high level of reciprocity in helpfulness between the
generations.

\vspace*{4ex}

\section{Conclusions}
\label{s_conclusions}

In this paper we have developed methods for analysing
intergenerational help and support that is exhanged between individuals
and their non-coresident parents. This involved specifying latent
variable models for data where the help given and received are measured
by multiple items for different types of help. The models include a
multivariate zero inflation component to allow for the fact that a large
proportion of people in our data gave or received no help of any kind.
Estimation of the models was done in a two-step fashion, where the
measurement model for the help items given latent helping tendencies was
estimated and fixed first, before the structural model for the joint
distibution of these latent variables given explanatory variables was
then estimated. The estimation of the structural model was carried out
using an MCMC algorithm implemented specifically for these models.

We analysed data from the UK Household Longitudinal Study, where the
respondents (the children in the parent-child dyads) are aged around
40 on average, and their parents around~70. The results of the analysis
indicated some strong predictors of helping, which may be interpreted in
terms of the capacities and needs of the two parties. The levels of help
given by the parents and by the children were positively correlated,
suggesting substantial cross-sectional reciprocity of help between the
generations. The survey data that we have used is extensive and rich in
many respects, but it also has some limitations. In particular, because
only one member of the dyad --- here the children --- were interviewed,
the data on their parents is limited. It would be preferable to survey
both parties directly, but this data collection design is difficult to
implement on a large scale.

The model proposed here is immediately applicable also to other
applications with the same structure, that is `doubly multivariate' data
with two sets of observed binary items measuring two latent variables.
For example, it could be used to analyse attitudes among couples, when
the interest was also on the concordance between the partners. Further,
the model could be extended in different ways, both for this and for
other applications. This would involve, in essence, combining the kinds
of structural models that would be used in each situation if the
variables of interest were directly observed with the kinds of
measurement models considered here when they are latent rather than
observed. For example, data where the dyads are grouped in natural
clusters could be accommodated in this way by including random effects
(higher-order latent variables) to allow for within-cluster
associations. An important instance of this is longitudinal data on
dyads, which will be needed for questions about levels and reciprocity
of intergenerational help over time. Models for longitudinal data can
also be specified in other ways; for example, \citet{steele+grundy21}
consider a dynamic (autoregressive) panel model that allows for unequal
spacing between the measurements, but simplifying the analysis in
another way by reducing giving and receiving help each to a binary
variable.

Another straightforward extension of the model is obtained by allowing
multiple latent variables for each member of the dyad, each measured by
their own multiple indicators. This would be needed, for example, if we
wanted to consider forms of financial and practical help separately from
each other. It is also possible for the same individuals to appear in
multiple dyads. For some such cases the structural model would be an
obvious extension of the models considered in this paper, for example if
we analysed data where survey respondents were asked about help that
they exchanged with their children as well as with their parents. In more
complex situations, such as for `round-robin' data where each individual
is paired with more than one other individual, the models should include
further role-specific latent variables for `actors' and
`partners' (or `givers' and `receivers'), as well as group (e.g.\
family) effects. This would define multivariate extensions of different
versions of the Social Relations Model (\citealt{kenny+lavoie84};
\citealt{snijders+kenny99}). \citet{ginetal20} have recently proposed
latent-variable formulations for such situations, and our measurement
models would add to them the element of zero inflation. These
combinations remain to be explored in future research.

\begin{table}[p]
\centering
\caption{Estimated parameters of the linear models for
the tendency to give help to ($\eta_{G}$) and to receive help
from ($\eta_{R}$) individuals' non-coresident parents, from
the estimated model for data from Wave 7 of the UK Household Longitudinal Study
described in Section \ref{ss_analysis_structural}.
The estimates are posterior means from MCMC samples (with posterior standard
deviations in parentheses).}

\vspace*{1ex}
\label{t_model_estimates}
\begin{tabular}{lrrcrr}
\hline
&
\multicolumn{2}{c}{Help to parents} &&
\multicolumn{2}{c}{Help from parents} \\
&  Estimate & (s.d.) & & Estimate & (s.d.)\\
\hline
\multicolumn{6}{l}{\rule{0pt}{3ex}\emph{Coefficients of explanatory variables
($\hat{\boldsymbol{\beta}}_{G}$ and $\hat{\boldsymbol{\beta}}_{R}$):}} \\[1ex]
Intercept &$-2.31^{***}$&(0.08)&&$-4.14^{***}$&(0.09)\\[.5ex]
\textbf{Respondent (child) characteristics}\hspace*{2em}         &  &&&         & \\
Age ($\times$10 years)          &
$0.04^{\phantom{***}}$&(0.04)&&$-0.55^{***}$& (0.06)\\[.5ex]
Gender                                 &&&          &           & \\
~~Female (vs.\ Male)            & $0.88^{***}$&(0.05)&& $0.72^{***}$ & (0.06)
\\[.5ex]
Partnership status                               &&&   &           & \\
~~Partnered (vs.\ Single)        & $-0.25^{***}$&(0.06)&&$-0.68^{***}$&(0.07)
\\[.5ex]
Employment status (vs.\ Employed)                        &&&           &           & \\
~~Unemployed                                        & $0.33^{***}$
&(0.13)&&$0.47^{***}$&(0.14) \\
~~Economically inactive                             &
$0.42^{***}$&(0.06)&&$0.25^{***}$&(0.08)
\\[.5ex]
\multicolumn{6}{l}{Age of youngest child in the respondent's own
household (vs.\ 0--1 years):}\\
~~No children                   &$0.20^{*\phantom{**}}$ &(0.10)&&$0.21^{**\phantom{*}}$&(0.10)\\
~~0--1 years &$0^{\phantom{***}}$ & && $0^{\phantom{***}}$ & \\
~~2--4 years                  &$-0.01^{\phantom{***}}$ &(0.12)&&$-0.00^{\phantom{***}}$&(0.10)\\
~~5--10 years                 &$0.21^{*\phantom{**}}$ &(0.11)&&$0.04^{\phantom{***}}$& (0.10)\\
~~11--16 years                 &$0.16^{\phantom{***}}$ &(0.12)&&$0.12^{\phantom{***}}$& (0.13)\\
~~$>16$ years                    &$0.21^{*\phantom{**}}$ &(0.12)&&$-0.12^{\phantom{***}}$& (0.15)\\[8pt]
\textbf{Parent characteristics}                     &   &&&        &
\\
Age of oldest parent ($\times$ 10 years)  &$0.32^{***}$
&(0.04)&&$0.04^{\phantom{***}}$& (0.05)\\[.5ex]
At least one parent lives alone (vs.\ No)  &$0.56^{***}$&(0.05)&
&$-0.32^{***}$& (0.06)\\[8pt]
\textbf{Child-parent characteristics} &&&&& \\
Travel time to nearest parent &&&&&\\
~~More than 1 hour (vs.\ 1 hour or less)
&$-1.05^{***}$ &(0.08)&&$-0.46^{***}$& (0.09)\\[3ex]
\emph{Residual variances (
$\hat{\sigma}^{2}_{G}$ and
$\hat{\sigma}^{2}_{R}$
):} &$2.13^{***}$&(0.09)&&$2.43^{***}$&(0.11)\\
\emph{Residual correlation ($\hat{\rho}_{GR}$):} &$0.50^{***}$ &(0.02) &&& \\
\hline
\multicolumn{6}{l}{{\footnotesize{The posterior credible interval excludes zero at
level 90\% (*), 95\% (**) or 99\% (***).}}}
\end{tabular}
\end{table}

\begin{table}[p]
\centering
\caption{Fitted probabilities of the zero-inflation latent
classes $(\xi_{G},\xi_{R})$, from the estimated model described in
Section \ref{ss_analysis_structural}, averaged over parameter values in
MCMC samples and over covariate values $\mathbf{X}_{i}$ in the sample of
dyads $i$ of a respondent and their parent(s), and odds ratios (OR) calculated from these averages. On the
first row the values of $\mathbf{X}_{i}$ are all as in the observed
data, while on the other rows one covariate is set to the same value for
every dyad as indicated, while the rest keep their sample values.
The last six columns show the average marginal probabilities of classes
$\xi_{G}=1$ and
$\xi_{R}=1$ (those who may give or receive help), their differences
between different covariate settings, and standard deviations of these
differences across the MCMC samples.}
\label{t_pi}

\vspace*{1ex}
\begin{small}
\begin{tabular}{lrrrrrcrrrrrr}\hline
&&&&&&&\multicolumn{6}{c}{Marginal probabilities}\\
Covariate &
\multicolumn{4}{c}{$p(\xi_{G}=j,\xi_{R}=k)$} & &
& \multicolumn{6}{c}{[with difference (and its SD)]} \\
setting &(0,0)&(0,1)&(1,0)&(1,1)&OR&&
\multicolumn{3}{c}{$p(\xi_{G}=1)$} &
\multicolumn{3}{c}{$p(\xi_{R}=1)$} \\
\hline
Sample &.17&.10&.15&.59&7.1 & &.73& &&.68& & \\[1ex]
\multicolumn{13}{l}{\textbf{Respondent (child) characteristics}}\\
\multicolumn{13}{l}{\emph{Age}}\\
\hspace*{1em}35 years&.13&.11&.10 &.66 &8.0&& .76&&&.77&&\\
\hspace*{1em}45 years&.22&.09&.13 &.56 &\hspace*{1em}10.4&&
.69&$-.08^{***}$&(.01)&.65&$-.12^{***}$&(.02)\\
\multicolumn{13}{l}{\emph{Gender}}\\
\hspace*{1em}Male&.19&.06&.14 &.61 &14.6&& .75&&&.67&&\\
\hspace*{1em}Female&.16&.12&.15 &.55 &4.7&&
.72&$-.03^{*\phantom{**}}$&(.02)&.69&$+.02^{\phantom{***}}$&(.02)\\
\multicolumn{13}{l}{\emph{Partnership status}}\\
\hspace*{1em}Single&.15&.11&.09 &.65 &9.3&& .74&&&.76&&\\
\hspace*{1em}Partnered&.18&.09&.16&.57 &6.7&&
.73&$-.01^{\phantom{***}}$&(.02)&.66&$-.10^{***}$&(.02)\\
\multicolumn{13}{l}{\emph{Employment status}}\\
\hspace*{1em}Employed&.16&.10&.14 &.60 &6.7&& .73&&&.70&&\\
\hspace*{1em}Unemployed&.20&.06&.20&.54 &8.6&&
.74&$+.00^{\phantom{***}}$&(.04)&.61&$-.09^{***}$&(.04)\\
\hspace*{1em}Inactive&.22&.07&.17&.54 &9.7&&
.71&$-.03^{\phantom{***}}$&(.02)&.61&$-.09^{***}$&(.02)\\
\multicolumn{13}{l}{\emph{Age of youngest child in the respondent's
own household}}\\
\hspace*{1em}No children&.20&.07&.14&.58 &11.9&&
.73&$-.02^{\phantom{***}}$&(.03)&.65&$-.12^{***}$&(.04)\\
\hspace*{1em}0--1 years&.12&.13&.11 &.64 &5.1&& .75&&&.77&&\\
\hspace*{1em}2--4 years&.11&.13&.10&.66 &5.3&&
.76&$+.01^{\phantom{***}}$&(.04)&.79&$+.02^{\phantom{***}}$&(.04)\\
\hspace*{1em}5--10 years&.14&.14&.12&.59 &4.7&&
.72&$-.04^{\phantom{***}}$&(.03)&.74&$-.04^{\phantom{***}}$&(.04)\\
\hspace*{1em}11--16 years&.22&.08&.20&.50 &7.4&&
.70&$-.05^{\phantom{***}}$&(.04)&.58&$-.20^{***}$&(.04)\\
\hspace*{1em}$>16$ years&.19&.07&.14&.60 &11.1&&
.74&$-.01^{\phantom{***}}$&(.04)&.67&$-.10^{**\phantom{*}}$&(.05)\\[1ex]
\multicolumn{13}{l}{\textbf{Parent characteristics}}\\
\multicolumn{13}{l}{\emph{Age of oldest parent}}\\
\hspace*{1em}70 years&.22&.10&.07 &.61 &20.0&& .67&&&.71&&\\
\hspace*{1em}80 years&.12&.11&.17&.60 &4.0&&
.77&$+.10^{***}$&(.01)&.70&$-.01^{\phantom{***}}$&(.02)\\
\multicolumn{13}{l}{\emph{At least one parent lives alone}}\\
\hspace*{1em}No&.20&.12&.12 &.56 &7.6&& .68&&&.68&&\\
\hspace*{1em}Yes&.13&.05&.17&.64 &9.6&&
.81&$+.13^{***}$&(.02)&.69&$+.01^{\phantom{***}}$&(.02)\\[1ex]
\multicolumn{13}{l}{\textbf{Child--parent characteristics}}\\
\multicolumn{13}{l}{\emph{Travel time to nearest parent}}\\
\hspace*{1em}1 hour or less&.15&.12&.17 &.57 &4.3&& .74&&&.69&&\\
\hspace*{1em}$>1$ hour&.23&.04&.09&.63 &37.9&&
.72&$-.02^{\phantom{***}}$&(.04)&.67&$-.01^{\phantom{***}}$&(.03)\\
\hline
\multicolumn{13}{l}{{\footnotesize{The posterior credible interval excludes zero at
level 90\% (*), 95\% (**) or 99\% (***).}}}
\end{tabular}
\end{small}
\end{table}

\textbf{Acknowledgements}

This research was supported by a UK Economic and Social Research Council
(ESRC) grant ``Methods for the Analysis of Longitudinal Dyadic Data with
an Application to Inter-generational Exchanges of Family Support'' (ref.
ES/P000118/1).

Acknowledgement of the data used in the paper:
Understanding Society (UKHLS) is an initiative funded by the Economic and Social
Research Council and various Government Departments, with scientific
leadership by the Institute for Social and Economic Research, University
of Essex, and survey delivery by NatCen Social Research and Kantar
Public. The research data are distributed by the UK Data Service.

\appendix
%\section{Appendices}
%\label{s_appendices}

\setcounter{equation}{0}
\renewcommand{\theequation}{A\arabic{equation}}

\section{Appendix: Details of the MCMC algorithm}
\label{s_appendices}

Here we describe the details of the tailored MCMC sampling algorithm for
the estimation of the structral model parameters $\boldsymbol{\psi}$
which was outlined in Section~\ref{ss_estimation_structural}. The
algorithm has been packed into an R (\citealt{r2020}) package.
[which will be available open source on an
author's GitHub page]. The main part of the algorithm was programmed in
{C\nolinebreak[4]\hspace{-.05em}\raisebox{.4ex}{\tiny\bf ++}}, where two
techniques are used to speed up the sampling procedure. First, for
sampling steps with non-standard distributions, adaptive rejection
sampling (\citealt{gilks+wild92}) is used, exploiting log-concavity of
the posterior density functions. This is used for two of the four main
steps of the algorithm, for sampling $\boldsymbol{\psi}_{\xi}$ and (some
of) $\boldsymbol{\eta}$, while the other two (sampling
$\boldsymbol{\psi}_{\eta}$ and $\boldsymbol{\xi}$) can be done very
efficiently from standard distributions. Second, parallel sampling
techniques, spread out across multiple processors, are used within each
MCMC iteration when there is no dependence between the quantities being
sampled. This can be done when sampling the latent variables
$\boldsymbol\zeta_i$ for different units $i$, and also when sampling the
two subsets of structural parameters $\boldsymbol\psi_{\xi}$ and
$\boldsymbol\psi_{\eta}$ separately from each other. This
parallelisation is implemented through OpenMP
{C\nolinebreak[4]\hspace{-.05em}\raisebox{.4ex}{\tiny\bf ++}} API
(\citealt{dagum+menon98}).

Let
$\boldsymbol\zeta^{(t)}=(\boldsymbol{\xi}^{(t)},\boldsymbol{\eta}^{(t)})$
and
$\boldsymbol\psi^{(t)}=(\boldsymbol{\psi}_{\xi}^{(t)},\boldsymbol{\psi}_{\eta}^{(t)})$
denote the values of the latent variables and the structural parameters
sampled in iteration $t=0,1,2,\dots$, where 0 denotes the initial
values. Given $\boldsymbol{\zeta}^{(t-1)}$, $\boldsymbol{\psi}^{(t-1)}$
and the observed data $(\mathbf{Y},\mathbf{X})$, the values for the next
iteration $t$ are sampled as follows:

\textbf{Imputation step}: Generating values for the latent variables $\boldsymbol{\zeta}$,
given the observed data and current values of the parameters
$\boldsymbol{\psi}$:

(1) Sampling $\boldsymbol{\xi}^{(t)}$ from
$p(\boldsymbol{\xi}|\boldsymbol{\eta}^{(t-1)},\mathbf{Y},\mathbf{X},\boldsymbol{\psi}^{(t-1)})$:
Draw  $\boldsymbol{\xi}_{i}^{(t)}=
(\xi_{Gi}^{(t)},\xi_{Ri}^{(t)})$ independently for $i=1,\dots,n$, from
multinomial distributions with probabilities
\begin{eqnarray}
\lefteqn{p(\xi_{G}=j,\xi_{R}=k|\boldsymbol{\eta}^{(t-1)},\mathbf{Y}_{i},\mathbf{X}_{i},\boldsymbol{\psi}^{(t-1)})}
\label{p_xi}
\\
&\propto&
p(\mathbf{Y}_{Gi}|\xi_{G}=j,\eta_{Gi}^{(t-1)},\mathbf{X}_{i};\tilde{\boldsymbol{\phi}}_{G})\,
p(\mathbf{Y}_{Ri}|\xi_{R}=k,\eta_{Ri}^{(t-1)},\mathbf{X}_{i};\tilde{\boldsymbol{\phi}}_{R})\,
p(\xi_{G}=j,\xi_{R}=k|\mathbf{X}_{i};\boldsymbol{\psi}_{\xi}^{(t-1)})
\nonumber
\end{eqnarray}
for $j,k=0,1$, where the structural model for $\boldsymbol{\xi}_{i}$
is specified by (\ref{structural2}), the measurement model is specified as in
(\ref{meas1})--(\ref{meas2}) for $\mathbf{Y}_{Gi}$ and similarly for
$\mathbf{Y}_{Ri}$, and
the parameters
of the measurement models are fixed at their estimated values
$\tilde{\boldsymbol{\phi}}_{G}$ and
$\tilde{\boldsymbol{\phi}}_{R}$ from the first step of the two-step
estimation (as described in Section \ref{ss_estimation_measurement})
throughout.
Note that here the probabilities
which involve $\xi_{G}=0$ are zero when $\mathbf{Y}_{Gi}\ne \mathbf{0}$,
and the ones which involve $\xi_{R}=0$ are zero when $\mathbf{Y}_{Ri}\ne
\mathbf{0}$. Conversely, when $\mathbf{Y}_{Gi}$ and/or $\mathbf{Y}_{Gi}$
is $\mathbf{0}$, the imputation assigns such a unit $i$
either to the corresponding zero-inflation class 0 or to class 1 for
the duration of iteration $t$.

(2) Sampling $\boldsymbol{\eta}^{(t)}$ from
$p(\boldsymbol{\eta}|\boldsymbol{\xi}^{(t)},\mathbf{Y},\mathbf{X},\boldsymbol{\psi}^{(t-1)})$:
Draw  $\boldsymbol{\eta}_{i}^{(t)}=
(\eta_{Gi}^{(t)},\eta_{Ri}^{(t)})$ independently for $i=1,\dots,n$, as
follows. First, draw $\eta_{Gi}^{(t)}$ from
\begin{equation}
p(\eta_{G}|\eta_{Ri}^{(t-1)},\boldsymbol{\xi}_{i}^{(t)},\mathbf{Y},\mathbf{X},\boldsymbol{\psi}^{(t-1)})
\propto
p(\mathbf{Y}_{Gi}|\xi_{Gi}^{(t)},\eta_{G},\mathbf{X}_{i};\tilde{\boldsymbol{\phi}}_{G})
\;
p(\eta_{G}|\eta_{Ri}^{(t-1)},\mathbf{X}_{i};\boldsymbol{\psi}_{\eta}^{(t-1)})
\label{p_etaG}
\end{equation}
and then $\eta_{Ri}^{(t)}$ from
\begin{equation}
p(\eta_{R}|\eta_{Gi}^{(t)},\boldsymbol{\xi}_{i}^{(t)},\mathbf{Y},\mathbf{X},\boldsymbol{\psi}^{(t-1)})
\propto
p(\mathbf{Y}_{Ri}|\xi_{Ri}^{(t)},\eta_{R},\mathbf{X}_{i};\tilde{\boldsymbol{\phi}}_{R})
\;
p(\eta_{R}|\eta_{Gi}^{(t)},\mathbf{X}_{i};\boldsymbol{\psi}_{\eta}^{(t-1)})
\label{p_etaR}
\end{equation}
where the conditional distributions for $\eta_{G}$ and $\eta_{R}$ on the
right-hand sides are the univariate normal distributions implied by
(\ref{structural1}). When $\xi_{Gi}^{(t)}=0$, in which case always
$\mathbf{Y}_{Gi}=\mathbf{0}$, the probability for $\mathbf{Y}_{Gi}$ in
(\ref{p_etaG}) is 1 by (\ref{meas1}), and $\eta_{Gi}^{(t)}$ is generated
directly from this normal distribution, whereas adaptive rejection
sampling is used when
$\xi_{Gi}^{(t)}=1$; the procedure for $\eta_{Ri}^{(t)}$ is
analogous, depending on whether or not $\mathbf{Y}_{Ri}$ is $\mathbf{0}$.

\textbf{Posterior step}: Drawing values for the model parameters
$\boldsymbol{\psi}$ from
their distributions given the observed data and current imputed values of the
latent variables $\boldsymbol{\zeta}$. These are standard posterior distributions of the
parameters of regression models for $\boldsymbol{\zeta}$ given $\mathbf{X}$, the bivariate linear model
(\ref{structural1}) for $\boldsymbol{\eta}$ and the multinomial
logistic model (\ref{structural2}) for $\boldsymbol{\xi}$. These
do not depend on each other, so these sampling steps can be carried out
in either order or in parallel.

(3) Sampling $\boldsymbol{\psi}_{\eta}$ from its posterior distribution
$p(\boldsymbol{\psi}_{\eta}|\mathbf{X},\boldsymbol{\eta}^{(t)}) \propto
p(\boldsymbol{\eta}^{(t)}|\mathbf{X};\boldsymbol{\psi}_{\eta})\,
p(\boldsymbol{\psi}_{\eta})$. These parameters are handled in two
blocks, $\boldsymbol{\beta}=(\boldsymbol{\beta}_{G}', \,
\boldsymbol{\beta}_{R}')'=\text{vec}(\mathbf{B})$ where
$\mathbf{B}=[\boldsymbol{\beta}_{G} \; \boldsymbol{\beta}_{R}]$, and
$(\sigma_{G}^{2}, \sigma^{2}_{R}, \rho_{GR})$ which define the
conditional  covariance matrix in (\ref{structural1}), which we denote
$\boldsymbol{\Sigma}_{\eta}$.
Here we define the notation specifically as
$\mathbf{X}=[\mathbf{X}_{1}\; \dots \; \mathbf{X}_{n}]'$,
$\boldsymbol{\eta}_{G}=(\eta_{G1},\dots,\eta_{Gn})'$,
$\boldsymbol{\eta}_{R}=(\eta_{G1},\dots,\eta_{Rn})'$ and
$\boldsymbol{\eta}=[\boldsymbol{\eta}_{G}\; \boldsymbol{\eta}_{R}]$. The
bivariate linear model (\ref{structural1}) can then be written as
$\text{vec}(\boldsymbol{\eta})\sim
N(\text{vec}(\mathbf{XB}),\boldsymbol{\Sigma}_{\eta}\otimes
\mathbf{I}_{n})$, where $\mathbf{I}_{n}$ denotes the $n\times n$
identity matrix.

We specify the prior distribution as
$p(\boldsymbol{\psi}_{\eta})=p(\boldsymbol{\beta})p(\boldsymbol{\Sigma}_{\eta})$,
where $p(\boldsymbol{\beta})\sim
N(\mathbf{0},\sigma^{2}_{\beta}\mathbf{I}_{2q})$ with
$\sigma^{2}_{\beta}=100$, and $p(\boldsymbol{\Sigma}_{\eta})\sim
\mathcal{W}^{-1}(\mathbf{I}_{2},2)$, an inverse Wishart prior for
$\boldsymbol{\Sigma}_{\eta}$. This is a `semi-conjugate' prior for
$\boldsymbol{\psi}_{\eta}$, meaning that conditional on
$\boldsymbol{\Sigma}_{\eta}$, the posterior distribution of
$\boldsymbol{\beta}$ is also multivariate normal, and conditional on
$\boldsymbol{\beta}$ the posterior of $\boldsymbol{\Sigma}_{\eta}$ is
inverse Wishart. Specifically, $\boldsymbol{\beta}^{(t)}$ is then
sampled from the distribution
\begin{equation}
p(\boldsymbol{\beta}|\boldsymbol{\eta}^{(t)},\mathbf{X},\boldsymbol{\Sigma}_{\eta}^{(t-1)})
\sim N(\boldsymbol{\mu}_{\beta}^{(t)},\mathbf{V}_{\beta}^{(t)})
\label{p_beta}
\end{equation}
where
\begin{eqnarray}
\mathbf{V}_{\beta}^{(t)} & = &
\left(
\mathbf{I}_{2q}/\sigma^{2}_{\beta} +
(\boldsymbol{\Sigma}_{\eta}^{(t-1)})^{-1} \otimes
(\mathbf{X}'\mathbf{X})\right)^{-1}
\hspace*{3em}\text{ and } \\
\boldsymbol{\mu}_{\beta}^{(t)} &=&
\mathbf{V}_{\beta}^{(t)}\left(
(\boldsymbol{\Sigma}_{\eta}^{(t-1)})^{-1}
\otimes
\mathbf{X}'\right)\text{vec}(\boldsymbol{\eta})
\label{p_beta_mu}
\end{eqnarray}
where $\otimes$ denotes the Kronecker product, and $\boldsymbol{\Sigma}_{\eta}^{(t)}$ is sampled from
\begin{equation}
p(\boldsymbol{\Sigma}_{\eta}|\boldsymbol{\beta}^{(t)},\mathbf{X},\boldsymbol{\eta}^{(t)})
\sim \mathcal{W}^{-1}(\mathbf{I}_{2} +
(\boldsymbol{\eta}^{(t)}-\mathbf{XB}^{(t)})'
(\boldsymbol{\eta}^{(t)}-\mathbf{XB}^{(t)}),n+2).
\label{p_Sigmaeta}
\end{equation}

(4) Sampling
$\boldsymbol{\psi}_{\xi}^{(t)}=\boldsymbol{\gamma}^{(t)}=(
\boldsymbol{\gamma}_{00}^{(t)\prime},
\boldsymbol{\gamma}_{01}^{(t)\prime},
\boldsymbol{\gamma}_{10}^{(t)\prime},
\boldsymbol{\gamma}_{11}^{(t)\prime})'$, where
$\boldsymbol{\gamma}_{00}^{(t)}=\mathbf{0}$,
from the posterior distribution
$p(\boldsymbol{\psi}_{\xi}|\mathbf{X},\boldsymbol{\xi}^{(t)}) \propto
p(\boldsymbol{\xi}^{(t)}|\mathbf{X};\boldsymbol{\psi}_{\xi})\,
p(\boldsymbol{\psi}_{\xi})$. This is done using conditional Gibbs
sampling, one parameter at a time.
We specify the prior distribution as $p(\boldsymbol{\psi}_{\xi})\sim
N(\mathbf{0},\sigma^{2}_{\gamma}\,\mathbf{I}_{3q})$, with
$\sigma^{2}_{\gamma}=100$. Letting $\gamma_{jkr}$ denote the $r$th
element of $\boldsymbol{\gamma}_{jk}$, we cycle over all $r=1\dots,q$ and over $(j,k)=(0,1), \; (1,0),\; (1,1)$
to draw $\gamma_{jkr}^{(t)}$ from
\begin{equation}
p(\gamma_{jkr}|\boldsymbol{\gamma}_{(jkr)}^{(t-1)},\mathbf{X},\boldsymbol{\xi}^{(t)})
\propto \left[
\prod_{i=1}^{n} \;
\frac{\prod_{u,v=0,1}\;
\exp(\boldsymbol{\gamma}_{uvr}^{(t-1)\prime}\mathbf{X}_{i})^{\delta_{iuv}^{(t)}}}{
\sum_{u,v=0,1}\; \exp(\boldsymbol{\gamma}_{uvr}^{(t-1)\prime}\mathbf{X}_{i})
}
\right]\; p(\gamma_{jkr})
\label{p_psiksi}
\end{equation}
where $\boldsymbol{\gamma}_{uvr}^{(t-1)}$ are vectors where
all the $\gamma$-parameters except for $\gamma_{jkr}$ are fixed at their most
recently sampled values (from iteration $t-1$ or $t$),
$\boldsymbol{\gamma}_{(jkr)}^{(t-1)}$ denotes all of these fixed
parameter values,
$\delta^{(t)}_{iuv}=I(\xi_{Gi}^{(t)}=u, \xi_{Ri}^{(t)}=v)$,
and $p(\gamma_{jkr})$ is the prior density of $\gamma_{jkr}$ implied by
$p(\boldsymbol{\psi}_{\xi})$, in our case $p(\gamma_{jkr})\sim N(0,100)$.
These $\gamma_{jkr}^{(t)}$ are generated using adaptive rejection sampling.

\bibliographystyle{chicago}

\end{document}